\documentclass[sigconf]{acmart}
\usepackage{graphicx}
\usepackage{amsmath}
\usepackage{booktabs}
\usepackage{algorithm}
\usepackage{algorithmic}
\usepackage{amsfonts}
\usepackage{multirow}
\usepackage{makecell}
\usepackage{subfigure}
\usepackage{color}
\usepackage{bm}
\usepackage{epstopdf}
\usepackage{url}
\usepackage[cal=cm]{mathalfa}
\usepackage{balance}
\usepackage{threeparttable}
\usepackage{lipsum}
\usepackage{enumitem}
 \usepackage{pifont}
 \usepackage{tabularx}
 \usepackage{makecell}
 \usepackage{float}
 \usepackage{tcolorbox}
\setlist{itemsep=0pt, topsep=5pt} %

\AtBeginDocument{%
  \providecommand\BibTeX{{%
    \normalfont B\kern-0.5em{\scshape i\kern-0.25em b}\kern-0.8em\TeX}}}

\renewcommand\footnotetextcopyrightpermission[1]{}

\begin{document}
\title{LLM-Alignment Live-Streaming Recommendation}

\author{Yueyang Liu$^*$, Jiangxia Cao$^*$, Shen Wang$^*$, Shuang Wen, Xiang Chen, Xiangyu Wu, Shuang Yang, Zhaojie Liu, Kun Gai, Guorui Zhou}
\thanks{* Equal contribution to this work.}
\affiliation{
  \institution{Kuaishou Technology, Beijing, China}
 \country{\{liuyueyang05, caojiangxia, wangshen, wenshuang, chenxiang08, wuxiangyu06, yangshuang08, zhaotianxing, zhouguorui\}@kuaishou.com} {gai.kun@qq.com}
}

\renewcommand{\shorttitle}{}

\begin{abstract}
In recent years, integrated short-video and live-streaming platforms have gained massive global adoption, offering dynamic content creation and consumption. Unlike pre-recorded short videos, live-streaming enables real-time interaction between authors and users, fostering deeper engagement. However, this dynamic nature introduces a critical challenge for recommendation systems (RecSys): the same live-streaming vastly different experiences depending on when a user watching.
To optimize recommendations, a RecSys must accurately interpret the real-time semantics of live content and align them with user preferences. Yet, in our industrial deployment, we observed that simply injecting LLM-generated live-streaming embeddings into the recommendation model yielded marginal improvements. We attribute this to a fundamental mismatch: while the LLM is trained on vision-language tasks, the RecSys learns from user-item interaction data. This embedding space misalignment limits the model’s effectiveness.

To bridge this gap, we propose LARM (LLM-Alignment Live-Streaming Recommendation), a novel framework that leverages multi-modal LLMs to address the temporal dynamics of live-streaming recommendations. LARM integrates three core innovations:
(1) We fine-tune a open-source LLM on live-streaming data to enhance its understanding of real-time content, enabling it to generate context-aware embeddings.
(2) We align LLM-generated embeddings with RecSys ID embeddings by projecting them into a unified space with a gaated mechanism, ensuring compatibility with user-item interaction signals.
(3) We transform aligned embeddings into learnable semantic code features, enriching the retrieval and ranking stages of the RecSys pipeline.
Extensive experiments demonstrate that LARM significantly outperforms baseline models, while ablation studies validate the necessity of each component in industrial settings.
\end{abstract}



\maketitle

\section{Introduction}

\begin{figure}[t!]
  \centering
  \includegraphics[width=8.5cm,height=6cm]{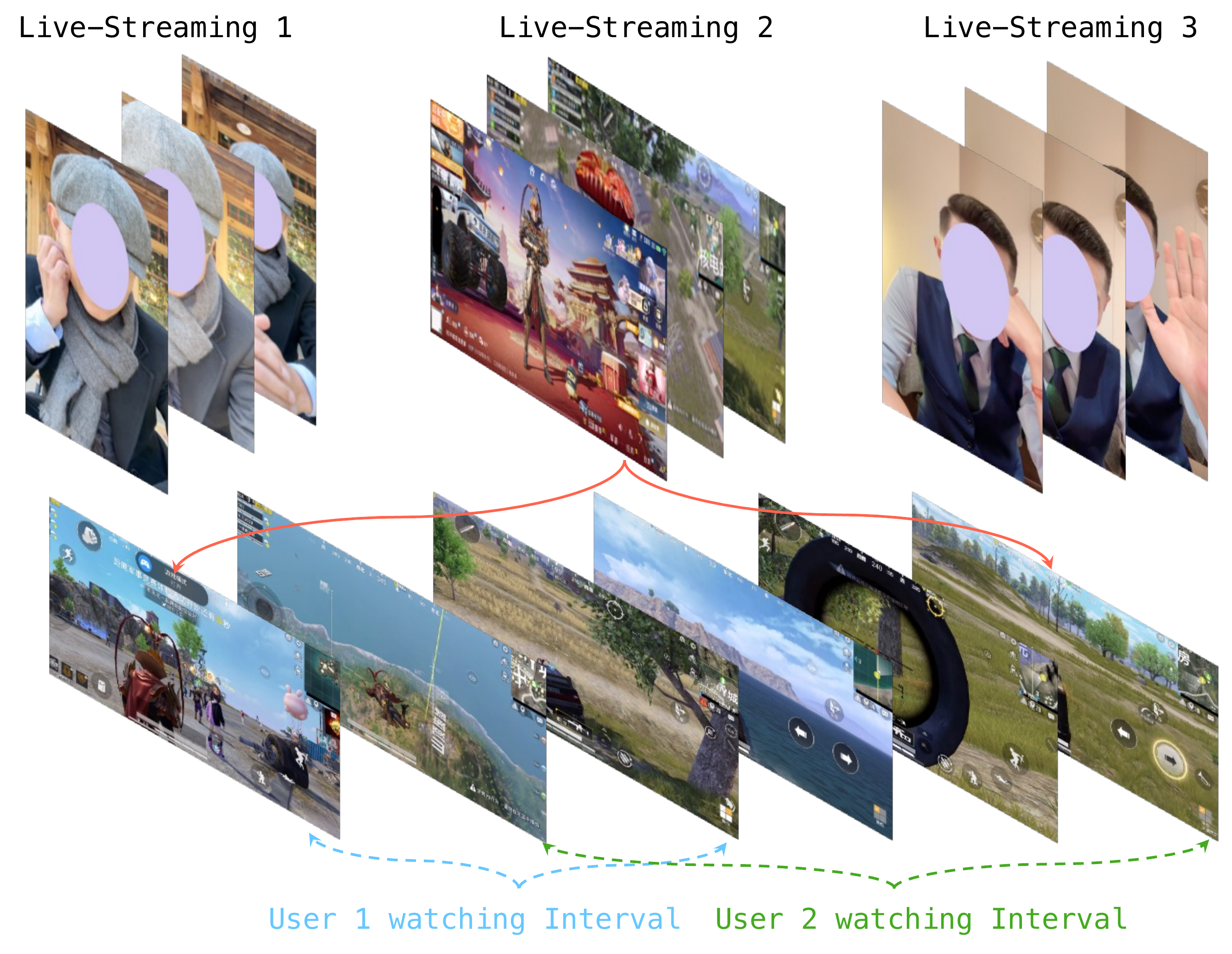}
  \caption{Two-types of conflicts: (i) authors' different live-streaming could have different topics; (ii) for a live-streaming, different users could watch different intervals.}
  \label{fig:intro}
\end{figure}

With the rapid advancement of information technology, multimedia content has gained increasing popularity online. 
Short-videos and live-streaming can convey significantly more information than traditional text and images, greatly enhancing user experience and attracting a worldwide audience, as demonstrated by popular platforms like TikTok, Kuaishou, and Xiaohongshu.

Among these medias, live-streaming is the most complex one, to connect the author and users to chat in real-time manner~\cite{crossmoment}.
Specifically, it can only be distributed in platform within the temporarily limited time when the author is living, and the content is ever changing and unpredictable.
Therefore, the distribution of live-streaming has a highly uncertain nature: \textit{for a same live-streaming, users will have different experience when watching live-streaming at different time point}.

Consequently, a challenging but valuable problem arises: How to reduce the recommendation uncertainty to distribute a live-streaming at the optimal moment to guarantee a better user experience?
A widely used solution to this issue involves gathering users' past interactions with positively viewed author IDs\footnote{Typically, author ID reflects their unique live-streaming style.}, and then recommending similar live streams or authors. Examples of such methods include DIN~\cite{din} and SIM~\cite{sim}.
However, in live-streaming services, this approach is not fine-grained enough to capture precise user interests for the following reasons:
\begin{itemize}[leftmargin=*,align=left]
    \item \textbf{Author-side conflict:} the learned author ID embedding aim to compress the overall information in author's past live-streamings. However, the author may cover different topics across various live-streamings.
    \item \textbf{User-side conflict:} in each live-streaming, users watch different live content duration, and some users even do not overlap at all.
\end{itemize}
To alleviate above two conflicts in Figure\ref{fig:intro}, we propose a potential way to record the certain semantics of corresponding live-streaming at users' watching sessions.
Once the model is able to comprehend the semantics of real-time live streaming, our RecSys can intelligently distribute live streams from other creators that match users' preferred semantics, thereby enhancing recommendation accuracy.

With the success of multi-modal large language model (LLMs) in global knowledge understanding, such as GPTs~\cite{gpt2}, Qwen~\cite{bai2023qwen}, DeepSeek~\cite{liu2024deepseek}, MiniCPM~\cite{yao2024minicpm}, many researchers have realized the potential of LLMs to enhance recommendation systems, such as NoteLLM~\cite{zhang2024notellm} and QARM~\cite{luo2024qarm} for note and short-video services.

In their designing architecture, considering the high computation cost of LLMs' inference, these works always follow a cascading cache paradigm to support industrial-scale RecSys:
\begin{itemize}[leftmargin=*,align=left]
    \item \textbf{LLM-RecSys knowledge tuning:} collecting a high-quality item-item corpus relationships from RecSys model to pre-train LLM to adapt RecSys Knowledge.
    \item \textbf{LLM Caching:} applying the pre-trained LLM to generate embeddings for each item and cache them.
    \item \textbf{RecSys Usage:} according to the LLM cache store, select related items' LLM embeddings to support recommendation model.
\end{itemize}
In this architecture, the downstream knowledge-aware LLM and RecSys are decoupled from each other, which enhances system efficiency in responding to large volumes of requests.

Following the above paradigm, we have explored several different ways to generate the item-item corpus to pre-train LLM and align it with RecSys knowledge in live-streaming scenarios. These approaches included using high-similarity author-author pairs from Swing~\cite{yang2020large}, authors gifted by the same users, and others. 
Unfortunately, these explorations ultimately proved unsuccessful for the following reasons:
\begin{itemize}[leftmargin=*,align=left]
    \item \textbf{Noisy live-streaming pairs:} although we can obtain the similarity between authors, it's still unclear which specific intervals of their live-streams share the most similarity.
    \item \textbf{Live-streaming context is too long for LLM:} unlike short videos that are just a few seconds long, live-streaming lasts for several hours, making it much more challenging to compress the speech, comments, and visuals of an entire stream.
\end{itemize}
As a result, the first step \textbf{LLM-RecSys knowledge tuning} is not an easy task in live-streaming scenario,  \textbf{is there another way to align the LLM embeddings with RecSys signals?}

In this paper, we propose a novel simple-yet-effective framework for \underline{L}ive-streaming \underline{A}lignment \underline{R}eco\underline{m}mendation, termed as \textbf{LARM}, which includes our first-hand practical experience on the multi-modal LLM deployment to support industrial RecSys.
In our roadmap, we focus on three technical points: 
\begin{enumerate}[leftmargin=*,align=left]
    \item How to tune a LLM for live-streaming services?
    \item How to align the LLM embedding with RecSys knowledge?
    \item How to leverage the aligned embeddings in RecSys model?
\end{enumerate}

\vspace{1em} 

\textbf{Domain-Specific LLM Tuning.} As we known, the open-source LLMs always have undergone two stages of training to enable it to have open question answering capabilities: pre-training~\cite{bert} and post-training (a.k.a, instruct-tuning)~\cite{lambert2024t}.
Building on this LLM foundation, for downstream usage, the model often requires fine-grained tuning to adapt to specific tasks and achieve better performance. For example, item-item multi-modal contrastive learning~\cite{clip} in QARM.
Since it is challenging to provide appropriate item-item pairs in live-streaming, we opt to construct high-quality question-answering tasks to help the LLM understand the live-streaming content.
Specifically, we first collect a million-scale live-streaming information in 30-second window, which includes images, speeches and comments).
Next, we design a sophisticated prompt to encourage a powerful 100B LLM to generate some question-answer pairs and reading comprehension results based on the live-streaming information.
Finally, according to the live-streaming multi-modal information and supervised tuning data, we fine-tune a 7B LLM to adapt to the corpus and inject live-streaming knowledge.
Once the 7B LLM is trained, we freeze it and use it to produce real-time live-streaming embedding every 30 seconds.

\vspace{1em} 

\textbf{Embedding Alignment.} Using the real-time live-streaming embeddings produced by the 7B LLM, we conducted a series of experiments with our RecSys model.
However, in our practice, we found that directly using the `RecSys knowledge un-aligned' LLM-generated real-time live-streaming embeddings as part of recommendation model inputs only resulted in slight performance gains.
The reason may be lie in the inconsistency of embedding spaces: the multi-modal LLM is supervised by a vision-language corpus, but recommendation model is supervised by user-item interaction logs, these separate spaces limit the model's performance.
Therefore, it is necessary to find a common space to align the knowledge from both RecSys and multi-modal knowledge.
Motivated by this, we devised a simple gate mechanism in the two-tower model to ensure that the author ID embedding and the LLM embedding are fused into the same space.
Though simple, we are surprisingly observed that the multi-modal embedding were not ignored in the learning process. The gate network assigning 0.65 weight for ID and 0.35 weight for LLM, which demonstrate that this learning method has not collapsed into the ID space, but rather found a new fusion space of RecSys and multi-modal knowledge.

\vspace{1em} 

\textbf{Embedding Quantification.} 
In downstream RecSys usage, to recommend live-streamings that align with users’ historical preferences and similar semantics, we need to store the real-time live-streaming semantics for each (user, live-streaming) pair at a given time.
However, storing the real-time high-dimensional LLM float embeddings in our user log collector is costly in terms of storage.
To alleviate the problem, we choose to quantify the gate-aligned author embeddings to Semantic IDs\cite{rqvae} to describe live-streaming status.
In our LARM, we map a 256 high-dimensional float embedding to 3 integer-based Semantic IDs, significantly reducing the required storage.
This approach allows us to enrich users' live-streaming viewing history with the corresponding interval semantics, enabling more intelligent and contextually relevant recommendations.

In summary, our contributions are as follows:
\begin{itemize}[leftmargin=*,align=left, itemsep=0pt]
    \item To the best of our knowledge, this paper is the first work to explore the alignment of RecSys and multi-modal knowledge in live-streaming scenario, which will shed light on other researchers to build more smart multi-modal RecSys.
    \item We propose a practical, simple and effective solution, LARM, which describes the overall roadmap of our industrial experiences in live-streaming deployment, including LLM tuning, RecSys-LLM alignment and downstream Usage. 
    \item We conduct extensive offline and online experiments to verify the effectiveness of LARM. Additionally, we present detailed case studies demonstrating that LARM successfully identifies a new fusion space that balances RecSys and multi-modal knowledge. Furthermore, LARM has been fully deployed in industrial live-streaming services.
\end{itemize}

\section{LARM Workflow}
In this section, we give our solution for the three technical points:
\begin{enumerate}[leftmargin=*,align=left]
    \item How to tune a LLM for live-streaming service?
    \item How to alignment the LLM embedding with RecSys knowledge?
    \item How to utilize the aligned embedding in RecSys model?
\end{enumerate}

\subsection{LLM Tuning for Live-Streaming}
To identify the ever-changing live-streaming content semantic, we consider using LLM to perform real-time embedding inference for each live-steaming.
However, real-time inference will bring huge computing pressure, thus we need to conduct trade-offs between the window size and the LLM parameter size.
In our practice, we select the 30-second window size to update live-streaming LLM embedding, and a 7B LLM for fast inference.

To encourage a 7B LLM to acquire knowledge in live-streaming, it typically requires further supervised fine-tuning with annotated live-streaming question-answering data. 
Particularly, it is time-consuming and inefficiency to annotate a large amount of data by human, thus we utilize an unsupervised method to collect our data corpus.
We first collect a million-scale live-streaming information in 30-second window (including author information, images, speeches and so on). 
Next, we write a sophisticated prompt to encourage a well-trained 100B LLM to generate some question-answer pairs and reading comprehensive results around the live-streaming information.
\begin{tcolorbox}[colback=blue!10!white,colframe=blue!50!black,title=Live-Streaming Instruction Generation Prompt]
\small
Live-streaming information: \#time, \#images, \#speeches, \#comments, \#author\_information;\\

You need to pay attention to the fact that the person in the live-streaming may not be the author himself, but someone else. \\

Please generate ten instruction questions as diverse as possible based on the provided live-streaming information. These questions are about facts or an understanding and evaluation of relevant content. Do not ask any questions that cannot be answered confidently. \\

Finally, you need to return the result in the following form:\\
1. \{"Question":..., "Answer":...\}\\
2. \{"Question":..., "Answer":...\}\\
\end{tcolorbox}
\begin{tcolorbox}[colback=blue!10!white,colframe=blue!50!black,title=Live-Streaming Reading Comprehension Prompt]
\small
Live-streaming information: \#time, \#images, \#speeches, \#comments, \#author\_information;\\

You need to pay attention to the fact that the person in the live-streaming may not be the author himself, but someone else. \\

Complete the following tasks: \\
1. Describe the current events and content in the live broadcast room in detail in one paragraph, including but not limited to:
Content: including detailed visual information, such as whether there is text, the look and feel of the picture, the tone, etc. \\
\#other\_tasks,\\

Finally, you need to return the result in the form of json:\\
\{ "Event and content": \{ \\
"Content": ..., \\
"Character": ..., \\
"Scene": ..., \\
"Event": ..., \\
"Atmosphere": ..., \\
"Highlights": ..., \\
"Target user group": ... \},\}\\
\end{tcolorbox}
In this way, we could employ the 100B LLM itself to generate a large amount of instructions according to our provided prompt guidelines and live-streaming information.
Based on collected instruction question-answer pairs and reading comprehension data, the 7B LLM is able to recognize the live-streaming background and make in-depth thinking based on itself world knowledge.
After tuning the 7B LLM, we could deploy it online to compress real-time live-streaming understanding results into an embedding and send it to the downstream RecSys.

\subsection{Gated Embedding Alignment}
For the `un-alignment' LLM embedding usage in downstream RecSys model, there are some other efforts to align them by inserting an additional contrastive objective between item-side ID embedding and LLM embedding, e.g., EM3~\cite{deng2024end}.
Although contrastive alignment paradigm attempts to find a common space to align them, we argue that it is relatively implicit and the learning pattern easily overfitting, since the LLM embedding and ID embedding are strictly one-to-one mappings.
In this section, we propose a more explicit gate alignment mechanism to enforce the alignmnet of ID and LLM embedding into a same space.

\begin{figure*}[t!]
  \centering
  \includegraphics[width=18cm,height=4.5cm]{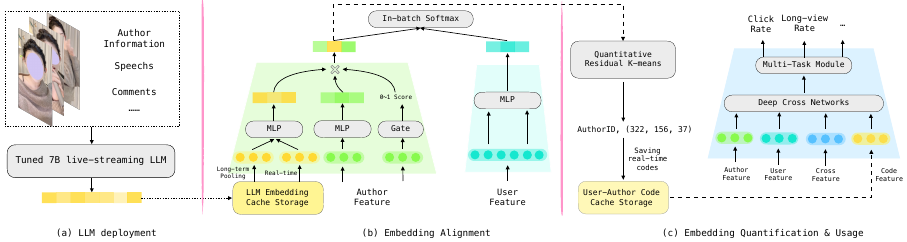}
  \caption{The workflow of LARM: (a) Tuning a LLM to produce real-time live-streaming embedding every 30s; (2) Gated fusion mechanism to align the RecSys ID space and multi-modal LLM space; (3) Quantifying the real-time aligned author embedding to record the certain semantics of corresponding live-streaming at users’ watching interval, to capture user interests accurately.}
  \label{fig:method}
\end{figure*}

As we know, the industry RecSys consists of two stages to make fast and precise prediction from a large amount items set.
(1) For fast, low-latency prediction, the first stage uses a two-tower paradigm-based \textbf{Retrieval model}~\cite{fm}, which disentangles the user-side and item-side information modeling to generate pure user and item representation. Thus we could forward the user-tower in inference stage to generate user representation to search small group relevant items from cached item representation store.
(2) For fine-grained, precise prediction, the second stage employs a multi-task paradigm-based \textbf{Ranking model}~\cite{home}, which mixes the user-side, item-side, and user-item-cross information into a complex, high-dimensional mixture representation. This model predicts  several probabilities aspects to select the best dozen items from the retrieved item candidates (e.g., click, long-view).

\subsubsection{Item-side Alignment}
Considering that the Ranking model focuses on user-item feature crossing and there is no module that can fully represent item-side information.
As a result, we conduct our gated alignment mechanism in the item-tower of Retrieval model to generate a multi-modal aware item representation, as shown in Figure~\ref{fig:method}(b).
For brevity, we utilize $\mathbf{aID} \in \mathbb{R}^{d}$ denote the live-streaming author ID embedding, $\mathbf{aLLM}_{30s} \in \mathbb{R}^{d}$ denote the real-time 30s embedding, and $\mathbf{aLLM}_{pooling} \in \mathbb{R}^{d}$ denote the long-term average pooling LLM embedding:
\begin{equation}
\small
\begin{split}
\mathbf{aLLMRep} = \texttt{MLP}_{llm}(&\mathbf{aLLM}_{30s}\oplus \mathbf{aLLM}_{pooling}),\\
\mathbf{aRecRep} = &\texttt{MLP}_{item}(\mathbf{aID}),\\
\lambda = \texttt{Sigmoid}&\big(\texttt{Gate}(\mathbf{aID})\big),\\
\mathbf{aFusionRep} = \lambda * \mathbf{aL}&\mathbf{LMRep} + (1 - \lambda) * \mathbf{aRecRep},
\end{split}
\label{finetune}
\end{equation}
where $\texttt{MLP}_{llm}(\cdot): \mathbb{R}^{2\times d}\xrightarrow{} \mathbb{R}^{d}$ and $\texttt{MLP}_{item}(\cdot): \mathbb{R}^{d}\xrightarrow{} \mathbb{R}^{d}$ are multi-layer FFN module, the gate network $\texttt{Gate}(\cdot): \mathbb{R}^{d}\xrightarrow{} \mathbb{R}$ is activated by the $\texttt{Sigmoid}$ function to generate a float importance value $\lambda \in (0, 1)$.
The $\mathbf{aFusionRep}$ is the final fused embedding merged the RecSys and LLM knowledge in the same space.
Next, we optimize the overall learning process by the user-item interaction data-streaming as follows:
\begin{equation}
\small
\begin{split}
\mathbf{uRecRep} = &\texttt{MLP}_{user}(\mathbf{uID}),\\
\mathcal{L}_{retrieval} = \texttt{In-batch-sof}&\texttt{tmax}(\mathbf{uRecRep}, \mathbf{aFusionRep}), \\
\end{split}
\label{finetune}
\end{equation}
where the $\mathbf{uID} \in \mathbb{R}^{d}$ denotes user ID embedding, and the $\mathcal{L}_{retrieval}$ is the training contrastive objective.
Although simple, we are surprisingly observed that the multi-modal embedding are not ignored in the learning process: the gate network assigning 0.65 weight for ID and 0.35 weight for LLM, which demonstrate that our gate mechanism has not collapsed into the ID space, but has found a new fusion space of RecSys and multi-modal knowledge.

\subsection{Real-Time Embedding Quantification}
Up to now, we have discussed how to obtain the fused representation of item-side live-streaming, denoted as $\mathbf{aFusionRep}$, which captures the real-time live-streaming RecSys and multi-modal information.
Meanwhile, user-side modeling plays a crucial role in recommendation systems. For instance, recommending other live-streamings based on the semantics of users' historically preferred live-streamings with similar content.
To achieve this, a straightforward approach would be to store the current multi-modal LLM embedding of each live-streaming when it is exposed to the user.
However, storing the high-dimensional float LLM embedding sequence for each user in the user log collector is prohibitively expensive.

\subsubsection{Live-streaming Semantic Codes}
To alleviate the problem, inspired by the RQ-VAE technique~\cite{rqvae} to quantify the embedding into several code IDs, we decide to transform the real-time LLM embedding into several codes to reduce the storage pressure.
Specifically, we first collect an \textbf{embedding corpus} of a -scale author embedding( $\mathbf{aFusionRep}$ ) as $\mathbf{A}$, and then apply a residual quantitative mechanism by K-means technique~\cite{luo2024qarm} to generate three-level codebooks with cluster number [512, 256, 128]\footnote{Different from billion-scale short-videos always with large cluster number, e.g., 8192. Live-streamings have only tens of millions of authors, so a smaller cluster number is sufficient.}:
\begin{equation}
\small
    \begin{split}
    \mathbf{C}^1 &= \texttt{Kmeans}(\mathbf{A}, 512),  \\
    \mathbf{C}^2 &= \texttt{Kmeans}(\mathbf{A}^1, 256), \mathbf{A}^1 = \mathbf{A}-\texttt{NearestRep}(\mathbf{A}, \mathbf{C}^1) \\
    \mathbf{C}^3 &= \texttt{Kmeans}(\mathbf{A}^2, 128),  \mathbf{A}^2 = \mathbf{A}^1-\texttt{NearestRep}(\mathbf{A}^1, \mathbf{C}^2)\\
    \end{split}
    \label{rqcodebook}
\end{equation}
where the $\mathbf{C}^1 \in \mathbb{R}^{512\times d}$, $\mathbf{C}^2 \in \mathbb{R}^{256\times d}$, $\mathbf{C}^3 \in \mathbb{R}^{128\times d}$ are the generated residual codebooks.
According to pre-trained codebooks, we could further quantify the real-time representation $\mathbf{aFusionRep}$ into three code IDs as:
\begin{equation}
\small
    \begin{split}
    \{c^1, c^2, c^3\} = \texttt{NearestCodeRep}(\mathbf{aFusionRep}, \mathbf{C}^1, \mathbf{C}^2, \mathbf{C}^3),
    \end{split}
    \label{codelist}
\end{equation}
In this way, we could compress the high-dimensional embedding as three integer codes, which is acceptable to \textbf{save them for each user/live-streaming pair at exposure moment} to enrich user-side information\footnote{In pratice, for 100 Million user with 10,000 historical sequence, LLM float-32bit embedding with 256 dimension is required 1024TB, while integer-8bit embedding with 3 dimension is required 3TB only.}.

\subsubsection{User-side Semantic Enriching}
Through accumulating the user/live-streaming pair Semantic IDs, we can restore the user's state when watching the live-streaming, to achieve more accurate user interest mining.
For example, in Ranking model\footnote{Actually, we have been utilized the Semantic codes in the user-tower of retrieval model. Since the author-side information decoupling of two-tower model, we apply the simple sum-pooling technique for user-side historical interacted live-streaming author sequence modeling.}, we could utilize the multi-modal aware Code IDs to empower the wide-used target-attention mechanism~\cite{din} as follows:
\begin{equation}
\small
    \begin{split}
    \textbf{Code}&\textbf{Feature} = \texttt{Attention}(Q = (\mathbf{aID}_1, \mathbf{c}^1, \mathbf{c}^2, \mathbf{c}^3), \\
        &K=\{(\mathbf{aID}_{u_1}, \mathbf{c}^1_{u_1}, \mathbf{c}^2_{u_1}, \mathbf{c}^3_{u_1}),\dots,(\mathbf{aID}_n, \mathbf{c}^1_{u_n}, \mathbf{c}^2_{u_n}, \mathbf{c}^3_{u_n})\}\\
        &V=\{(\mathbf{aID}_{u_1}, \mathbf{c}^1_{u_1}, \mathbf{c}^2_{u_1}, \mathbf{c}^3_{u_1}),\dots,(\mathbf{aID}_n, \mathbf{c}^1_{u_n}, \mathbf{c}^2_{u_n}, \mathbf{c}^3_{u_n})\}\\
    \end{split}
    \label{codelistesu}
\end{equation}
where Attention query is the $(\mathbf{aID}_1, \mathbf{c}^1, \mathbf{c}^2, \mathbf{c}^3)$ indicates the target live-streaming author ID and its real-time Code IDs embedding,
and key/value are the users historical interacted live-streaming authors and the \textbf{Semantic Code IDs the users watched}, e.g., $\{(\mathbf{aID}_{u_1}, \mathbf{c}^1_{u_1}, \mathbf{c}^2_{u_1}, \mathbf{c}^3_{u_1}),\dots$$,(\mathbf{aID}_n, \mathbf{c}^1_{u_n}, \mathbf{c}^2_{u_n}, \mathbf{c}^3_{u_n})\}$, the $u_n$ denotes sequence length.
To optimize the component, we could directly minimize the cross-entropy binary classification loss in multi-task manner:
\begin{equation}
\small
\begin{split}
\hat{y}^{ctr}, \hat{y}^{lvtr}, ... = \texttt{multi-task}([&\mathbf{aID}, \mathbf{uID}, \mathbf{CrossFeature}, \mathbf{CodeFeature}])\\
\mathcal{L}_{ranking} = - \sum_{xtr}^{{\texttt{ctr}, \dots}} \big(y^{xtr}&\texttt{log}{(\hat{y}^{\texttt{xtr}})} + (1-y^{xtr})\texttt{log}{(1-\hat{y}^{xtr}})\big)
\end{split}
\label{modeling}
\end{equation}
where the $ \texttt{multi-task}(\cdot)$ is a mixture-of-expert network, the $\hat{y}^{ctr}, \hat{y}^{lvtr}$ are the predicted user/live-streaming pair click rate, long-view rate, etc.
the $y^{ctr}, y^{lvtr}$ are ground-truth user/live-streaming labels, and $\mathcal{L}_{ranking}$ is the training objective in ranking model.

\section{Experiments}

In this section, we conduct a series of detailed experiments on our industrial live-streaming recommendation system (RecSys) to evaluate the effectiveness of our proposed LARM.

The primary goal is to answer the following research questions:

\begin{itemize}[leftmargin=*,align=left]
\item \textbf{RQ1}: How does LARM improve the performance of Retrieval and Ranking models in an offline setting?
\item \textbf{RQ2}: What impact does LARM have on online performance in real-world A/B tests?
\item \textbf{RQ3}: Which groups of authors are most impacted by the application of LARM?
\item \textbf{RQ4}: Does the gated alignment mechanism in LARM lead to better embeddings?
\item \textbf{RQ5}: Does our quantification mechanism enable meaningful divisions among authors?
\end{itemize}

\subsection{Experimental Setup}
\subsubsection{Data-streaming} 
We conducted offline experiments with LARM in both the retrieval and ranking stage at industrial live streaming scenario.
\textbf{It is worth noting that our platform includes hundreds of million-scale of activate users, million-scale live-streaming authors, and 3 billions of user/lives-streaming interactions every day}.

\subsubsection{Retrieval Metrics}
In the retrieval stage, we follow the wide-used user-to-item paradigm: using the generated user tower embedding $\mathbf{uRecRep}$ to search the nearest items from the cached $\mathbf{aFusionRep}$ embedding storage.
In offline evaluation, we simulate the online retrieval requests results to estimate different model ability, e.g., we calculate the \textbf{Hit Rate@1000} between the real user viewed items whether they are retrieved in the the nearest 1000 searched items of models, which defined as follows:
\[
\texttt{Hit Rate}_{u}@1000 = \frac{|\mathcal{P} \cap \mathcal{R}|}{|\mathcal{R}|}
\]
\text{Where:}
\begin{itemize}[leftmargin=*,align=left]
    \item \( \mathcal{P} \) represents the set of authors user $u$ watched.
    \item \( \mathcal{R} \) represents the nearest 1000 of retrieved authors for user $u$.
\end{itemize}

\begin{table}[t]
    \centering
    \caption{Offline Results (\%) of Hit Rate in Retrieval Stage.}
    \begin{tabular}{lc}
    \Xhline{0.8pt}
        \textbf{Models} & \textbf{Hit Rate@1000} \\ \Xhline{0.4pt}
        Two-Tower & 13.86 \\ 
        $\quad$Replaced aID Emb by LLM Emb & 7.65 \\ \hline
        $\quad$+ LARM author-side Gated Fusion & 17.03 \\ 
        \multirow{1}{*}{\makecell[c]{$\quad\quad$+ LARM user-side Semantic Codes}} & \multirow{1}{*}{17.80} \\  \Xhline{0.8pt}
    \end{tabular}
\label{exp:retrieval}
\end{table}

\subsubsection{Ranking Metrics}
In the ranking stage, we introduce LARM Semantic Codes to model users' live-streaming interests. For each user, we build a sequence of their 50 most recent valid views (watch time $\geq$ 3s), using the corresponding LARM Semantic Codes as multi-modal representations.
In offline evaluation, we apply the binary point-wise metrics AUC (ROC version) and GAUC (AUC
weighted by the exposures of each user) to reflect the model's predictive ability, which defined as follows:
\begin{equation}
\begin{split}
\texttt{GAUC} = \sum_{u} w_u \texttt{AUC}_u \quad \texttt{where}\ \ w_u = \frac{\texttt{sample}_u}{\texttt{all samples}},
\end{split}
\label{gauc}
\end{equation}
where the $w_u$ denotes the user $u$'s sample ratio.
Besides, the Ranking stage always has multiple prediction objective, here we show the click rate (CTR), Long-view rate (LVTR), Effective-view rate (ETR), Like rate (LTR), Comment rate (CMTR), and Gift rate (GTR).

\begin{table*}[t!]
\centering
\caption{Offline Results(\%) in term AUC of GAUC in Live-Streaming Rank Model.}
\begin{tabular}{lcccccccccccc}
\Xhline{0.8pt} %
\multirow{2}{*}{\textbf{Model}} & \multicolumn{2}{c}{\textbf{Click}} & \multicolumn{2}{c}{\textbf{Long-view}} & \multicolumn{2}{c}{\textbf{Like}} & \multicolumn{2}{c}{\textbf{Comment}} & \multicolumn{2}{c}{\textbf{Gift}} \\ 
\cline{2-11} %
                       & \textbf{AUC} & \textbf{GAUC} & \textbf{AUC} & \textbf{GAUC} & \textbf{AUC} & \textbf{GAUC} & \textbf{AUC} & \textbf{GAUC} & \textbf{AUC} & \textbf{GAUC} \\ 
\Xhline{0.4pt} %
Ranking Model          & 0.844 & 0.640 & 0.839 & 0.654 & 0.935 & 0.693 & 0.944 & 0.713 & 0.976 & 0.716 \\
$\quad$+LLM user-side Semantic Codes & +0.04 & +0.04 & +0.08 & +0.06 & +0.03 & +0.02 & +0.04 & +0.01 & +0.03 & +0.08 \\
$\quad$+LARM user-side Semantic Codes & +0.18 & +0.2 & +0.31 & +0.35 & +0.24 & +0.17 & +0.30 & +0.12 & +0.17 & +0.21 \\ 
\Xhline{0.8pt} %
\end{tabular}
\label{exp:Rank_offline_results}
\end{table*}

\begin{table*}[t!]
\centering
\caption{Online LARM A/B Testing Performance (\%) of Industrial Live-Streaming Services}
\setlength{\tabcolsep}{13pt}{
\begin{tabular}{cccccccc}
\Xhline{0.8pt} %
\multirow{2}{*}{\textbf{Applications}} & \multirow{2}{*}{\textbf{Stages}} & \multicolumn{3}{c}{\textbf{Core Metrics}} & \multicolumn{3}{c}{\textbf{Interact Metrics}} \\ \cline{3-8}
                              &                         & \textbf{Click} & \textbf{Watch Time} & \textbf{Gift Count} & \textbf{Like}       & \textbf{Comment}     & \textbf{Follow}    \\ 
\Xhline{0.4pt} 
\multirow{2}{*}{Scenario\#1}     & Retrieval                  & +0.144     & +0.144     & +1.119     & +0.340     & +0.256      & +0.450    \\
                              & Rank                    & +0.296     & +0.072      & +0.622     & +2.517     & +1.790      & +0.083    \\ \hline
\multirow{2}{*}{Scenario\#2}& Retrieval                  & +0.106     & +0.050      & -0.043     & -0.872     & -0.495      & +1.087    \\
                              & Rank                    & +0.486     & +0.170      & +0.679     & +2.795     & +1.335      & +0.625    \\ 
\Xhline{0.8pt} %
\end{tabular}
}
\label{tab:Online_results}
\end{table*}

\subsection{Offline Performance (RQ1)}

\subsubsection{Retrieval Stage Performance}
To accurately estimate the offline contribution of LARM in the retrieval stage, we replay the user's real online requests and calculate the Hit Rate@1000: for each live-streaming viewed by the user, whether it is retrieved by the corresponding model in the top 1000 results.
As shown in Table~\ref{exp:retrieval}, here we conduct several model variants on the wide-used two-tower models:
\begin{itemize}[leftmargin=*,align=left]
\item The `Replaced aID Emb by LLM Emb' indicates that we mask the ID-based author embedding while only utilizing the real-time LLM embedding to represent each author.
\item The `+ LARM author-side Gated Fusion' means utilizes our gated alignment mechanism at item-tower.
\item The `+ LARM user-side Semantic Codes' further enrich the user-tower by the quantified semantic codes.
\end{itemize}
According to Table~\ref{exp:retrieval}, we have the following observations:
(1) Directly removing the ID-based author embedding, the prediction accuracy largely declined, which demonstrates only rely the multi-modal information is difficult to build a powerful RecSys.
(2) Equipped with our gated mechanism, the two-tower model prediction performance achieves significant improvement (13.86 $\xrightarrow{}$17.03), which verifies our LARM does not collapsed into the ID space, but has found a
new fusion space of RecSys and multi-modal knowledge.
(3) Enriching the user-side watching semantic information, the prediction performance could further enhanced, which reveals 
 that utilizing semantic is a promising way to solve the ID conflict problem in live-streaming.

\subsubsection{Ranking Stage Performance}
For the ranking stage, we directly give the offline AUC and GAUC improvements across all engagement metrics including clicks, likes, and comments, as shown in Table~\ref{exp:Rank_offline_results}.
Specifically, here we conduct two ranking model variants:
\begin{itemize}[leftmargin=*,align=left]
\item The `+LLM user-side Semantic Codes' means that we directly quantify the un-alignment LLM embedding to Semantic codes to enrich user-side watching history.
\item The `+LARM user-side Semantic Codes' utilizes the our gated alignment embedding quantified Semantic codes to enrich user-side watching history.
\end{itemize}
According to Table~\ref{exp:Rank_offline_results}, we have the following observations:
(1) By quantifying the original LLM embedding, the prediction accuracy can be improved, which reveals that adding semantic features to the user’s watching history is neccesary for live-streaming interests modeling.
(2) Compared to the LLM semantic codes, our gated alignment Semantic codes yields a large improvement further, which prove aligning the RecSys and LLM embedding into a same space is vital for multi-modal information modeling.

\begin{figure}[t!]
  \centering
  \includegraphics[width=7cm,height=9cm]{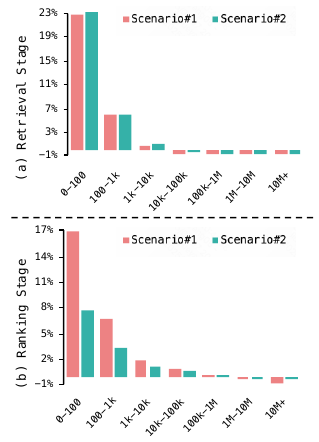}
  \caption{Online exposure changes across author group with different amount fans.}
  \label{fig:exposure}
\end{figure}

\subsection{Online A/B Tests (RQ2)}
To estimate the real performance gains achieved by our LARM framework, we deployed our LARM to serve the online retrieval and ranking models in our online A/B testing system at Scenario\#1 and Scenario\#1 (two major live-streaming services in our platform). 
Model performance was evaluated using core metrics including watch time, effective view, gift count, and click, along with interaction metrics such as likes and comments. 
The experimental results (Table~\ref{tab:Online_results}) demonstrate that LARM generates significant positive effects on user penetration while simultaneously improving both viewing duration and user engagement behaviors. 
These improvements were consistently observed across both platform variants, confirming the robustness of our approach. The findings highlight LARM's effectiveness in optimizing for both platform business objectives and user experience metrics.

\begin{figure*}[t!]
  \centering
  \includegraphics[width=16cm,height=10cm]{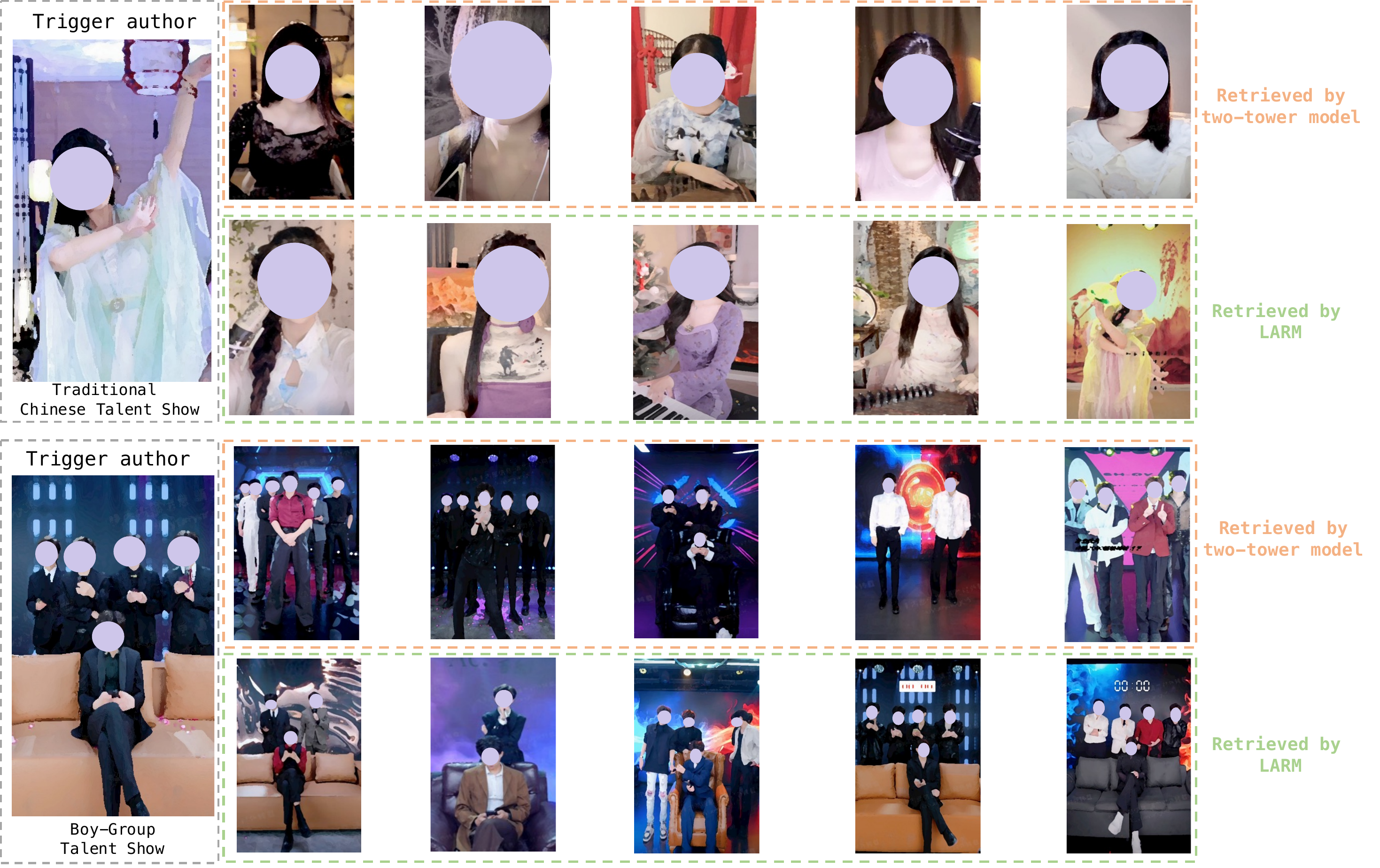}
  \caption{Different models' author-to-author retrieved author results for the left trigger author.}
  \label{fig:a2a}
\end{figure*}

\subsection{Author Group Analysis (RQ3)}
In LARM, we have two-steps iteration to enhance the ID-based RecSys: the first-step is applying a gate mechanism to align the author-side RecSys embedding and LLM embedding into a same space, the second-step is utilizing the Semantic code to enrich user-side history information.
Interestingly, we found that the average score of gate weights is changed, the first-step is 0.35 and the second-step is 0.47, which validates the cross-side mutual reinforcement between user and author multi-modal information.
Therefore, we were curious about what difference that LARM would make to our live-streaming RecSys.
To answer the question, we divide all authors into 7 groups according to their followed fans, e.g., 0-100, 100-1,000 to the 10,000,000+.
For these author groups, we further draw a bar chart to show the changes in exposure of the author ecosystem with our LARM (exposure means a successful distribution to our users).
As shown in Figure~\ref{fig:exposure}, we could observe that: with the number of fans increases, the number of exposures our LARM gives actually decreases in both of Retrieval and Ranking stage.
Such phenomenon validates our LARM  is not sensitive to whether the author are popular or not, while tend to support the long-tailed author group to build a fair environment for our live-streaming services.

\subsection{Author-Author Visualization (RQ4)}
In this section, we aim at answering the following question: does our gated alignment mechanism in LARM lead to better embeddings?
Intuitively, the ideal fused embedding should reflect the RecSys ID-based characters and the multi-modal world-knowledge characters simultaneously.
To explore it, we first deploy author-to-author retrieval service which caches all the authors' embeddings, and next input some trigger authors to service to obtain the nearest searched other authors.
Here we give two real searched cases in Figure~\ref{fig:a2a}, the left is the trigger authors, while the right is the searched results of ID-based Two-Tower model and our LARM.
From them, we have the following conclusions:
(1) For the first traditional Chinese talent show author, the Two-Tower model retrieved many female authors but ignore the live-streaming style. Compared with it, the other authors retrieved by our LARM are much better, and basically all of them match the style of the trigger author.
(2) For the second boy group talent show author, the Two-Tower model actually retrieved other boy group talent show authors. However, our LARM not only retrieved them, but also captured the `sofa-posed vision signals'.
According to them, we could safely claim that our gate mechanism has not collapsed into the ID space, but has found a
new fusion space of RecSys and multi-modal knowledge.

\begin{figure}[t!]
  \centering
  \includegraphics[width=8cm,height=4cm]{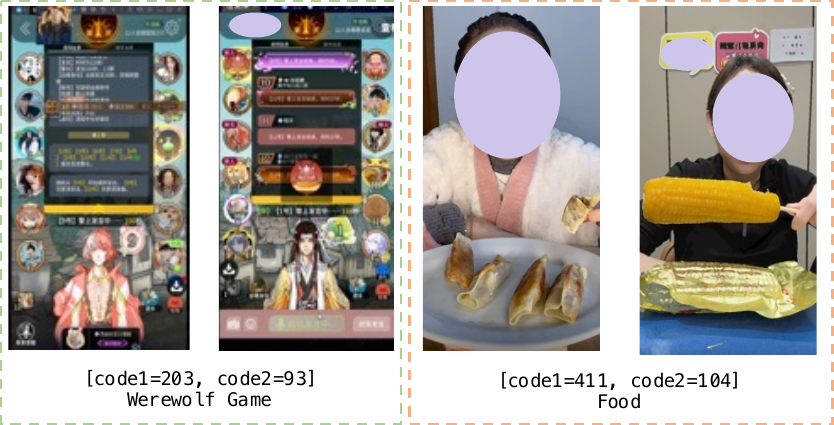}
  \caption{Case Study of LARM quantifying codes.}
  \label{fig:code}
\end{figure}

\subsection{Quantification Code Visualization (RQ5)}

In Section 2.3, we implement RQ-VAE quantization on the embeddings generated by LARM through cross-level encoding to obtain hierarchical semantic codes. To validate whether these codes effectively capture streamers' high-level semantic representations within the LARM framework, we perform visual analysis of code distributions. Our analysis reveals distinct semantic clustering within shared primary-secondary code pairs. 
Here we show some cases in Figure~\ref{fig:code}, we surprise find that: when the first two-level Semantic Code of authors are same, they are often in the same category!
In our practice, the Semantic code patterns: [203,93] represents werewolf-themed gaming content, while the [411,104] corresponds to the food authors.
This empirically validates that LARM successfully synthesizes multimodal and user behavioral signals to learn higher-order semantics, demonstrating its capability to capture nuanced content taxonomies through integrated signal processing.

\section{Related Works}
In this section, we briefly review the public researches of live-streaming and LLM for recommendation techniques.

\subsection{Live-Streaming Recommendation}
To our knowledge, the Twitch (a game live-streaming platform) proposed some early works for live-streaming recommendation, for instance, the LiveRec~\cite{rappaz2021recommendation} considers a special characteristic that user will watching a live-streaming author multiple times, and employ attention mechanism to capture such watching frequency behaviour pattern.
Besides, for the ultra long-view characteristic in live-streaming service (e.g., watching more than 1 hour), they further proposes a re-weight technique to increase the sample weight with user watching time~\cite{chen2022weighing}.
Although simple, they has been proven to be a widely proven and effective solution.
In addition to a pure live-streaming platforms, the live-streaming will also be integrated in other platforms, e.g., Taobao (a shopping platform), TikTok (short-video platform) and so on.
To exploit the users' multi-service behaviour, the eLiveRec~\cite{zhang2023cross} uses other interaction behaviours in product/short-video domain, and FARM~\cite{li2025farm} further introduce a cross-attention mechanism between different domain item sequence.
In addition to the above characteristics, live-streaming also has some high-light moments, at which the user experience is often significantly improved.
Therefore, some works are focuses on detect such high-light moments adaptively, such as the ContentCTR~\cite{deng2023contentctr} utilizes the modality information and the LiveForesighter~\cite{lu2025liveforesighter} models the periodicity of author behavior through the idea of time series.
Furthermore, instead of the designing of the model structure, some works have been iteratively the designing on the recommendation architecture for live-streaming services.
The Moment\&Cross~\cite{crossmoment} proposes a real-time live-streaming data-streaming and first-only label-mask learning framework, and the Sliver~\cite{liang2024ensure} gives a re-recommendation mechanism to alleviate the exposure delay problem.

\subsection{LLM for Recommendation}
In recent years, the LLM evolution has far exceeded our expectations, and has repeatedly refreshed the SOTA records and having human-like understanding capabilities, e.g., ChatGPTs~\cite{gpt2}, Qwen~\cite{bai2023qwen}, DeepSeek~\cite{liu2024deepseek} for text understanding, DALL-E, KLing for image/video generation.
In recommendation area, a hot topic is how to unlock the power of LLM to achieve a more smart recommendation~\cite{sheng2024enhancing, simtier}.
Actually, in industry, there are two different evolution directions in the past years to empower RecSys by LLM:
(1) Utilizing LLM learning paradigm to train recommendation models, (2) Using the pre-trained LLM to provide the world-knowledge for RecSys.

For the first branch methods, the KuaiFormer~\cite{liu2024kuaiformer} applies Transformers as encoder backbone in Retrieval stage, achieving SOTA performances than other multi-interests methods.
The HUST~\cite{hstu} with M-Falcon techniques stacks Transformers~\cite{bert} as encoder backbone in Ranking stage, reaching a "one sample, multi item" learning paradigm.
The MARM~\cite{lv2024marmunlockingfuturerecommendation} transfers the KV cache idea to Ranking model designing, successfully implemented a multi-layer Transformer structure with linear computation-cost growth.
The OneRec~\cite{deng2025onerec} apply the reinforcement learning paradigm to iterate a Transformer-based Retrieval model, which enables the Ranking model influence in the Retrieval model, achieving session-level recommendations.

For the second branch methods, the M3CSR~\cite{chen2024multi} fuse the learnable multi-modal tag ID embedding and freezing multi-modal LLM embedding to combine the recommendation and multi-modal knowledge.
the EM3~\cite{deng2024end} conducts the contrastive learning between the item ID and its multi-modal embedding to align them at shopping recommendation scenarios.
The NoteLLM~\cite{zhang2024notellm} compresses the item modality information into a embedding, and then utilize a high-quality item-item pair from recommendation model to align the multi-modal and recommendation knowledge.
The QARM~\cite{luo2024qarm} further quantify LLM embeddings as Semantic IDs, which alleviates the representation unlearning problem in downstream usage.

\section{Conclusion}

In this paper, we propose a novel LLM embedding and RecSys embedding alignment method, LARM, to enhance the LLM effectiveness for live-streaming recommendation.
Specifically, we first share our experience that how to tune LLM for live-streaming multi-modal information understanding, and used it to produce live-streaming LLM embedding in real-time manner.
According to the LLM embedding, we then employs a gating mechanism to regulate the proportional integration of the ID-based RecSys knowledge and multi-modal world knowledge. 
In our experiments, we find the average gate value is 0.35, which effectively demonstrates our LARM could made a trade-off between behavior-aware signals and multi-modal signals.
Furthermore, we also quantify the alignment embedding as semantic codes to empowering the users' historical live-stream sequences, to achieve a more smart recommendation to distribute live-streamings with users’ preferred semantics.
Empirically experimental results on the industrial data-streaming validates our LARM effectiveness in the Retrieval and Ranking stages.
Detailed ablation studies and case studies demonstrate our LARM is a promising way to adapt the LLM knowledge into RecSys models.

\balance
\bibliographystyle{ACM-Reference-Format}
\bibliography{sample-base-extend.bib}
\end{document}